\documentclass[a4paper,10pt]{article}
\usepackage{mathptmx}
\usepackage{amsmath}
\usepackage{amssymb}
\usepackage[english]{babel}
\date{}
\title{\textsc{Gravitational Radiation in Noncommutative Gravity}}
\author{A. Jahan, N. Sadegnezhad\\
Research Institute for Astronomy and Astrophysics of Maragha
(RIAAM)\\ Maragha, IRAN, P. O. Box: 55134 - 441\\
jahan@riaam.ac.ir}
\begin{document}
\maketitle
\begin{abstract}
The gravitational radiation power of a binary system in a noncommutative space is derived and it's rate of the period decrease is calculated to first order in noncommutativity parameter. By comparing the theoretical results with the observational data of the binary pulsar PSR 1913+16, we find a bound on the noncommutativity parameter.\\\
PACS: 11. 10. Nx, 04. 20. -q, 04. 30. -w\\\
\end{abstract}

\section{Introduction}
The need to cure the infinities plaguing the quantum fields was the first motivation to enlarge the Lorentz symmetry to include the noncommutative algebra [1, 2]. However, in recent decade interests in noncommutative theories gained a considerable attention because of the discoveries in string theory which imply that noncommutative space-time may be an inherent part of the high energy (Planck scale) physics. A space is noncommutative if it's coordinates satisfy
\begin{equation}\label{1}
[\widehat{x}^\alpha,\widehat{x}^\beta]=i\theta^{\alpha\beta},\quad\quad\theta^{\alpha\beta}=-\theta^{\beta\alpha}.
\end{equation}
 As a result of "$\theta$-deformation" of the algebra of space-time coordinates, one must replace the usual product among the fields with Weyl-Moyal product or $\star$-product [2], i.e.
\begin{equation}\label{1}
\phi_1(\widehat x)\star\phi_2(\widehat x)\equiv \lim_{x\rightarrow y}e^{\frac{i}{2}\theta^{\mu\nu}\partial^x_\mu\partial^y_\nu}\phi_1(x)\phi_2(y).
\end{equation}
From the Feynman's rules point of view, the only effect of the $\star$-product is to modify the $n$-point interaction vertices (3$\leq n$) by the phase factor [2]
\begin{equation}\label{1}
\tau(p_1,\ldots,p_n)=e^{-\frac{i}{2}\sum^n_{a<b}\,p_a\wedge p_b}.
\end{equation}
where $p_a\wedge p_b=\theta_{\mu\nu}p^\mu_a p^\nu_b$. Here the momentum flow of the $a$-th field into the vertex is denoted by $p_a$. In the case of gravitational field the effect of noncommutating coordinates is to modify the Newton potential as [3-6]
\begin{equation}\label{1}
V_\theta=-G\frac{m_am_b}{r}-G\frac{m_am_b}{2r^3}\,\textbf{L}\cdot\boldsymbol{\theta}+O(\theta^2).
\end{equation}
where $\textbf{L}$ is the particle's angular momentum. The possible effects of the $\theta$-deformed gravitational potential (4) on the celestial dynamics is considered by several authors [3-5].\\\
In present work we shall consider the gravitational radiation of a two body system in a circular motion with deformed potential between the massive bodies. We follow a quantum field theoretic approach to derive the potential (4) in next section. In section 3, we will consider the two body problem and re-derive the deformed potential using the so-called Bopp shift. In section 4, the gravitational radiation power and the period decrease rate are calculated for the system up to first order in noncommutativty parameter $\theta$. Finally, in section 5, the period decay of the model is compared with the  observational data of binary pulsar PSR 1913+16 to obtain a bound on the noncommutativty parameter.
\section{Deformed Newton Potential}
In a NC flat background the interaction between gravitational and scalar fields is [6, 7]
\begin{equation}\label{1}
\mathcal{L}_{int}=-\frac{\gamma}{2}{h}^{\mu\nu}\star
\Big[\partial_\mu{\phi}\star\partial_\nu{\phi}-\frac{1}{2}\big(\partial_\alpha{\phi}\star\partial^\alpha{\phi}-m^2\phi\star\phi\big)\Big],\quad \gamma^2=32\pi G.
\end{equation}
Therefore, with the aid of formula (3) we obtain the deformed momentum space 2 scalar-1 graviton vertex factor as
\begin{equation}\label{1}
\tau_{\alpha\beta}^\theta(p,p')=-\frac{i\gamma}{2}(p_\alpha p'_\beta+p'_\alpha p_\beta-\eta_{\alpha\beta}p\cdot p')e^{\frac{i}{2}\textbf{p}\wedge
\textbf{p}'}.
\end{equation}
where we have assumed $\theta_{\mu 0}=0$ implying $\theta^{ij}p^i q^j\rightarrow\textbf{p}\wedge \textbf{q}$ to avoid the problematic features of the noncommutative models [2]. Now, let us look at a typical two-body scattering mediated by a graviton. For the spinless particles with masses $m_a$ and $m_b$ the scattering amplitude is [6]
\begin{eqnarray}\label{8}
\mathcal M_\theta&=&\tau^{\mu\nu}_\theta(p_a,p'_a)D_{\mu\nu,\alpha\beta}(p_a-p'_a)\tau^{\alpha\beta}_\theta(p_b,p'_b),\\\nonumber
&=&\frac{4\pi G}{(p_a-p'_a)^2}\bigg(\Big[(p_a+p_b)^2-m^2_a-m^2_b\Big]^2+\Big[(p_a-p'_b)^2-m^2_a-m^2_b\Big]^2\\\nonumber
&-&\Big[(p'_a-p_a)^2+4m^2_am^2_b\Big]^2\bigg)e^{i \textbf{p}\wedge
(\textbf{p}- \textbf{p}')},
\end{eqnarray}
where the momentum-space graviton propagator is
\begin{equation}\label{1}
D_{\mu\nu\alpha\beta}(q)=-\frac{i}{2q^2}(\eta_{\mu\alpha}\eta_{\nu\beta}+\eta_{\mu\beta}\eta_{\nu\alpha}-\eta_{\mu\nu}\eta_{\alpha\beta}).
\end{equation}
In the non-relativistic limit we have
\begin{subequations}
\begin{align}
(p'_a-p_a)^2&\approx - \textbf{q}^{\,2},\\
(p_a+p_b)^2&\approx(m_a+m_b)^2,\\
(p_a-p'_b)^2&\approx(m_a-m_b)^2+\textbf{q}^{\,2}.
\end{align}
\end{subequations}
By substituting (9) in (7) we find the deformed gravitational potential
\begin{eqnarray}\label{8}
U_{\theta}(\textbf{x})&=&-\frac{1}{4m_am_b}\int\frac{d^3q}{(2\pi)^3}\mathcal M_{\theta}(\textbf{q})e^{i\textbf{q}\cdot \textbf{x}},\\\nonumber
&=&-G\frac{m_am_b}{\sqrt {({x^i-\frac{1}{2}\theta^{ij}p^j})({x^i-\frac{1}{2}\theta^{ik}p^k })}},\\\nonumber
&=&-\frac{\kappa}{r}-\frac{\kappa}{2r^3}\,\textbf{L}\cdot\boldsymbol{\theta}+O(\theta^2).
\end{eqnarray}
with $\kappa=Gm_am_b$.
\section{Two body Problem in NC Gravity}
The hamiltonian describing two particles $a$ and $b$ interacting via the Newton potential is
\begin{equation}\label{1}
H=\frac{\textbf{p}^2_a}{2m_a}+\frac{\textbf{p}^2_b}{2m_b}-\frac{\kappa}{|\widehat{\textbf{x}}_a-\widehat{\textbf{x}}_b|}.
\end{equation}
The classical canonical structure of the above system in NC space has the form
\begin{subequations}
\begin{align}
\{\widehat{x}_a^i,\widehat{x}_a^j\}&=\{\widehat{x}_b^i,\widehat{x}_b^j\}=\theta^{ij},\\\
\{\widehat{x}_a^i,\widehat{p}_a^j\}&=\{\widehat{x}_b^i,\widehat{p}_b^j\}=\delta^{ij}.
\end{align}
\end{subequations}
where we have used the correspondence $\frac{1}{i}[A,B]\rightarrow\{A,B\}$ to achieve the classical canonical structure from its quantum counterpart [3, 4]. We introduce the new set of coordinates
\begin{subequations}
\begin{align}
\textbf{X}&=\textbf{x}_{a}-\textbf{x}_b,\\\
\textbf{X}_c&=\frac{m_a\textbf{x}_{a}+m_b\textbf{x}_b}{m_a+m_b},
\end{align}
\end{subequations}
to rewrite (11) as
\begin{equation}\label{1}
H=\frac{\textbf{p}^2_c}{2(m_a+m_b)}+\frac{\textbf{p}^2_{X}}{2\mu}-\frac{\kappa}{|\widehat{\textbf{X}}|},
\end{equation}
with classical canonical structure given by
\begin{subequations}
\begin{align}
\{\widehat{X}^i,\widehat{X}^j\}&=2\theta^{ij},\\\
\{\widehat{X}^i,\widehat{P}^j_X\}&=\delta^{ij}.
\end{align}
\end{subequations}
Now, the so-called Bopp shift, i.e.
\begin{equation}\label{1}
\widehat{X}^i\rightarrow X^i=\widehat{X}^i+\theta^{ij}P^j_X,
\end{equation}
allows one to introduce the variable which fulfills the standard canonical structure, i.e. $\{{X}^i,{X}^j\}=0$.
By assuming that the center of mass is fixed, i.e. $\textbf{p}_c=0$, and on substituting $\theta\rightarrow 2\theta$ (c.f. (15.a)) the deformed Hamiltonian becomes
\begin{equation}\label{1}
H_\theta=\frac{\textbf{p}^2_{X}}{2\mu}-\frac{\kappa}{R}-\frac{\kappa}{R^3}\,\textbf{L}\cdot\boldsymbol{\theta},\quad {R=|\textbf{X}}|.
\end{equation}
For a circular motion i.e. $\dot R=0$ the equation of motion yields
\begin{equation}\label{1}
\frac{\partial H_\theta}{\partial R}=\mu R^3{\omega^2_\theta}-\kappa{\mu\theta\omega_\theta\cos\alpha}-{\kappa}=0.
\end{equation}
where we have used $L=\mu R^2\omega_0$ with $\omega^2_0=G\frac{m_a+m_b}{R^3}=\frac{\kappa}{\mu R^3}$. Here $\alpha$ denotes the angle between $\boldsymbol{\theta}$ and $\textbf{L}$. From (18) one finds the angular velocity of the system as
\begin{eqnarray}
{\omega_\theta}&=&\frac{1}{2}\omega^2_0{\mu\theta\cos\alpha}+\sqrt{\omega^2_0+\frac{1}{4}\omega^4_0{\mu^2\theta^2\cos^2\alpha}},\\\nonumber
 &=&\omega_0+\frac{1}{2}\omega^2_0{\mu\theta\cos\alpha}+O(\theta^2).
\end{eqnarray}
\section{Period Decay in a Compact Binary}
For a binary system located at $X^3=0$ plane, the coordinates of the bodies $a$ and $b$ circulating around the center of mass, are
\begin{subequations}
\begin{align}
 X^1_a &=\frac{\mu}{m_a} R\cos\omega_\theta t,\quad X^1_b =-\frac{\mu}{m_b} R\cos\omega_\theta t,\\
 X^2_a &=\frac{\mu}{m_a} R\sin\omega_\theta t,\quad  X^2_b=-\frac{\mu}{m_b} R\sin\omega_\theta t,\\
 X^3_a &=  X^3_b =0.
\end{align}
\end{subequations}
The total gravitational power radiated by the system is $P=\frac{G}{45c^5}\langle\dddot{D}^{ij} \dddot{D}^{ij} \rangle$ where the quadrupole moment is [8]
\begin{subequations}
\begin{align}
D^{11}&=\mu R^2(3\cos^2\omega_\theta t -1),\\\
D^{22}&=\mu R^2(3\sin^2\omega_\theta t -1),\\\
D^{21}&=D^{12}=3\mu R^2\sin\omega_\theta t\cos\omega_\theta t,
\end{align}
\end{subequations}
The radiated power by the both particles, $P_\theta=P_{\theta a}+P_{\theta b}$, becomes
\begin{eqnarray}\label{1}
P_\theta=-\frac{dE_\theta}{dt}&=&\frac{32}{5c^5}G\mu^2R^4\omega^6_\theta,\\\nonumber
&\simeq&\frac{32}{5c^5}G\mu^2R^4\omega^6_0+\frac{96}{5c^5}G\mu^3R^4\omega^7_0\theta\cos\alpha.
\end{eqnarray}
The energy of system is
\begin{eqnarray}
E_\theta&=&\frac{1}{2}\mu\omega_\theta^2R^2-\frac{\kappa}{R}-\frac{\kappa}{R}\mu\omega_0\theta\cos\alpha,\\\nonumber
&\simeq&\frac{1}{2}\mu\omega_0^2R^2-\frac{\kappa}{R}+\frac{1}{2}\mu^2\omega_0^3R^2\theta\cos\alpha-\frac{\kappa}{R}\mu\omega_0\theta\cos\alpha,\\\nonumber
&=&-\frac{\kappa}{2R}-\frac{1}{2}\sqrt{\frac{\mu\kappa^3}{R^5}}\theta\cos\alpha.
\end{eqnarray}
Therefore the rate of energy lose takes the form
\begin{equation}\label{1}
-\frac{dE_\theta}{dt}=-\frac{\kappa}{2R^2}\Big(1+\frac{5}{2}\omega_0\theta\mu\cos\alpha\Big)\frac{dR}{dt}.
\end{equation}
Thus, by equating the left hand side of (22) with (24), one obtains
\begin{equation}\label{1}
\frac{dR}{dt}=-\frac{64}{5c^5}\Big(\frac{G}{R}\Big)^3{m_am_b(m_a+m_b)}\,\frac{1+3\omega_0{\mu\theta\cos\alpha}}{1+\frac{5}{2}\omega_0\mu\theta\cos\alpha}.
\end{equation}
For $\theta=0$ the above expression coincides with the well-known textbook result [8]. From (25) and by virtue of $\tau\equiv\dot T=3\pi\sqrt{\frac{\mu R}{\kappa}}\dot R$ we obtain the rate of period decay as
\begin{eqnarray}\label{1}
\tau_\theta&=&-\frac{192}{5}\frac{\pi m_am_b}{(m_a+m_b)^{\frac{1}{3}}}\Big(\frac{2\pi G}{T}\Big)^{\frac{5}{3}}\frac{1+3\omega_0{\mu\theta\cos\alpha}}{1+\frac{5}{2}\omega_0\mu\theta\cos\alpha},\\\nonumber
&\simeq&\tau_0+\Delta\tau.
\end{eqnarray}
where $\Delta\tau=\frac{1}{2}\tau_0\omega_0\mu\theta\cos\alpha$. Again, for $\theta=0$ we are left with the standard result for the rate of (circular) orbit decay [9].
\section{PSR 1913+16 Binary System}
The masses of the pulsar and its companion in PSR 1913+16 binary system are
\begin{subequations}
\begin{align}
m_p &= 1.44 \,M_\odot,\\
m_c &= 1.38 \,M_\odot.
\end{align}
\end{subequations}
and the eccentricity of system is $e=0.61$. For non-circular orbit, i.e. for $e\neq 0$ case, the orbit decay rate $\tau_0$ includes the factor $f(e)$, which satisfies $f(0.61)=11.85$ [9]. The theoretical and observed values for the orbit decay rates, the reduced mass and period of the system, respectively are [9]
\begin{subequations}
\begin{align}
\tau_0&=-2.42\times 10^{-12}\,\textrm{sec/sec},\\\
\tau_0^{obs}&=-2.40  \times 10^{-12}\, \textrm{sec/sec},\\\
\mu&=0.7\,M_\odot=1.39\times 10^{30}\, \textrm{kg},\\\
T&=27898.56\ \textrm{sec}.
\end{align}
\end{subequations}
So, by assuming $\cos\alpha=1$, from the constraint
\begin{equation}\label{1}
\Delta \tau < \big|\tau_0-\tau_0^{obs}\big|,
\end{equation}
we find
\begin{equation}\label{1}
\theta  <1.6323\times10^{-29}.
\end{equation}
One must note that the above result for the noncommutativity parameter is valid as an estimation since the left hand side of constraint (30) does not include the factor accounting for the eccentricity of the system due to the fact that our analysis is restricted to the circular orbit.

\end{document}